\documentclass[prl,twocolumn,reprint,preprintnumbers,nofootinbib,superscriptaddress]{revtex4-2}
\input{header.tex}
\preprint{}
\usepackage{xcolor}

\begin{document}

\reportnum{-2}{CERN-TH-2024-159}

\title{Primordial neutrinos and new physics: novel approach to solving neutrino Boltzmann equation}

\author{Maksym Ovchynnikov}
\email{maksym.ovchynnikov@cern.ch}
\affiliation{Theoretical Physics Department, CERN, 1211 Geneva 23, Switzerland}
\affiliation{Institut für Astroteilchenphysik, Karlsruher Institut für Technologie, Germany}

\author{Vsevolod Syvolap}
\email{v.syvolap@umail.leidenuniv.nl}
\affiliation{Instituut-Lorentz, Leiden University, The Netherlands}

\date{\today}

\begin{abstract}
Understanding how new physics influences the dynamics of cosmic neutrinos during their decoupling is crucial in light of upcoming precise cosmological observations and the need to reconcile cosmological and laboratory probes. Existing approaches to solving the neutrino Boltzmann equation are often model-dependent, computationally inefficient, and yield contradictory results. To solve this problem, we introduce a novel method to comprehensively study neutrino dynamics. We apply this method to several case studies, resolving the discrepancy in the literature about the impact of non-thermal neutrinos on $N_{\rm eff}$ and providing important insights about the role of decaying new physics particles on MeV plasma. 
\end{abstract}

\maketitle

\textbf{Introduction.} Future observations of the Cosmic Microwave Background (CMB) by experiments such as the Simons Observatory~\cite{SimonsObservatory:2018koc} are poised to achieve unprecedented precision in measuring cosmological parameters. Notably, they aim for a percent-level accuracy in the determination of the effective number of relativistic neutrino species, $N_{\text{eff}}$, which tells us how much primordial neutrinos contributed to the energy density of the plasma at the times when they were ultrarelativistic~\cite{ParticleDataGroup:2024cfk}. This enhanced precision opens a window to probe new physics that may have existed at MeV temperatures, which could have significantly influenced neutrino properties in the Early Universe. Such new physics may not only alter the value of $N_{\text{eff}}$ but also modify the shape of the neutrino energy spectrum, which is essential for knowing the dynamics of the primordial neutrons (being a seed for Big Bang Nucleosynthesis~\cite{Sabti:2020yrt}), and the cosmological impact of neutrino mass~\cite{Alvey:2021sji}.

A systematic investigation into how new physics influences neutrinos becomes essential in this light. The central point is an efficient, model-agnostic method for solving the Boltzmann equation for the evolution of the neutrino distribution function. Currently, such an approach is absent, resulting in incoherent advancements in the field. Moreover, existing state-of-the-art studies yield qualitatively contradictory results for the class of new physics models that inject high-energy neutrinos at MeV temperatures~\cite{Dolgov:2000jw,Boyarsky:2021yoh,Mastrototaro:2021wzl,Rasmussen:2021kbf}. In this Letter, we introduce a novel method that not only allows us to draw significant model-independent conclusions but also enables the exploration of specific physical scenarios that existing methods cannot address. Our companion paper~\cite{Ovchynnikov:2024rfu} provides a detailed description of this approach and validates it against current state-of-the-art methods. Here, we summarize the main findings and present case studies delivering important physics insights.

\begin{figure}[t!]
    \centering
    \includegraphics[width=0.45\textwidth]{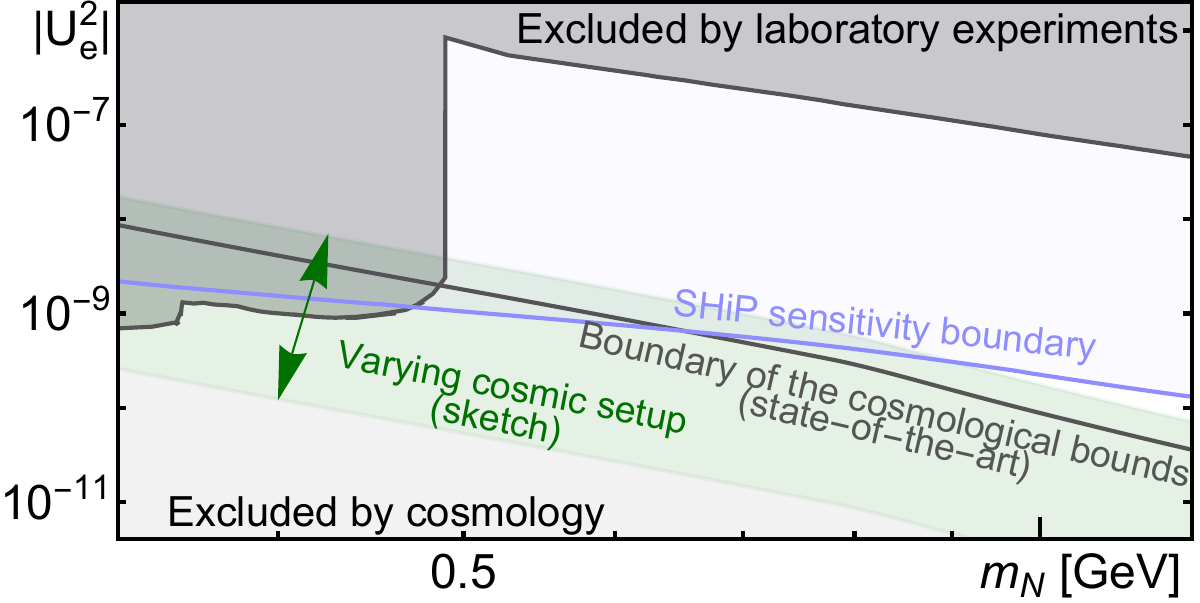}
    \caption{Parameter space of Heavy Neutral Leptons~\cite{Asaka:2005an,Asaka:2005pn} (HNLs) in the plane of their mass $m_N$ and the modulus of the coupling $|U_e|^2$. The past cosmological (the gray line shows the state-of-the-art calculations from~\cite{Sabti:2020yrt,Boyarsky:2021yoh}) and laboratory searches constrain complementary ranges of $|U_{e}|^{2}$, which is essential for defining the target for future laboratory experiments like SHiP~\cite{Aberle:2839677}. However, the past cosmological studies lack recent theoretical advances in the dynamics of the HNL decay products in the primordial plasma~\cite{Akita:2024ork,Akita:2024ta} as well as fix the minimal cosmological setup instead of varying it. By addressing these points, the line actually becomes the uncertainty bound (sketched by the green region). Due to the computational complexity, there are no existing approaches that robustly study this question. It is possible with the new approach to solve the neutrino Boltzmann equation presented in this paper.}
    \label{fig:hnl-complementarity}
\end{figure}

\textbf{New physics and neutrinos.} Various well-motivated new physics models have been proposed that can affect primordial neutrinos. These can be broadly classified into \emph{non-standard neutrino interactions}~\cite{Archidiacono:2013dua,Forastieri:2015paa,deSalas:2021aeh,Berryman:2022hds}, where neutrinos interact with new light mediators, altering their decoupling and thermal history; scenarios involving a \emph{neutrino-antineutrino asymmetry}~\cite{Dolgov:2002ab,Wong:2002fa,Grohs:2016cuu,EscuderoAbenza:2020cmq,Gelmini:2020ekg,Froustey:2021azz,Escudero:2022okz,Froustey:2024mgf}, leading to differences in the number densities of neutrinos and antineutrinos; and models introducing \emph{decaying new physics particles}, often referred to as long-lived particles (LLPs)~\cite{Dolgov:2000jw,Hannestad:2004px,Kanzaki:2007pd,Ruchayskiy:2012si,Fradette:2017sdd,Fradette:2018hhl,Hasegawa:2019jsa,EscuderoAbenza:2020cmq,Sabti:2020yrt,Gelmini:2020ekg,Boyarsky:2020dzc,Boyarsky:2021yoh,Mastrototaro:2021wzl,Rasmussen:2021kbf}. These particles may have been copiously produced in the Early Universe and then survived until MeV temperatures, injecting energy into the neutrino and electromagnetic sectors. Examples of LLPs are~\cite{Alekhin:2015byh} Heavy Neutral Leptons, axion-like particles, dark photons, $B-L$ mediators, Higgs-like scalars, neutralinos, generic particles participating in late reheating scenarios, and many others.

The cosmological probes of LLPs are of special importance, because they are complementary to laboratory searches, with the former exploring the domain of large lifetimes $\tau \gtrsim 10^{-2}\text{ s}$~\cite{Boyarsky:2020dzc} and the latter the domain of small lifetimes. The complementarity helps us to define the target parameter space for future experiments~\cite{Beacham:2019nyx,Antel:2023hkf,Aberle:2839677}, see Fig.~\ref{fig:hnl-complementarity}. Because of this, it is necessary to robustly understand the cosmological constraints, and, in particular, the impact of new physics on cosmic neutrinos at MeV temperatures, which defines the boundary of the constraints.

\textbf{Challenges and problems.} The target question requires solving the neutrino Boltzmann equation:
\begin{equation}
    \frac{\partial f_{\nu_{\alpha}}}{\partial t} - Hp\frac{\partial f_{\nu_{\alpha}}}{\partial p} = \mathcal{I}_{\rm coll}[f_{
    \nu_{\alpha}},p]
\end{equation}
Here, $f_{\nu_{\alpha}}(p,t)$ is the neutrino distribution function in the momentum space, $p$ is the momentum, $H \equiv \dot{a}/a$ is the Hubble parameter, and $\mathcal{I}_{\rm coll}$ is the collision integral, containing information about the interactions of neutrinos and source terms.

Several challenges severely restrict existing studies in the field. \emph{First}, there is extreme complexity in solving the neutrino Boltzmann equation. The most accurate way so far is to consider the approach of the discretization of the comoving momentum $y = p\cdot a$, firstly developed in~\cite{Hannestad:1995rs} for study standard evolution of neutrinos and improved since then~\cite{Grohs:2015tfy,Hasegawa:2019jsa,Gariazzo:2019gyi,Akita:2020szl,Froustey:2020mcq,Drewes:2024wbw}. It has been applied to study new physics particles~\cite{Dolgov:2000jw,Hannestad:2004px,Sabti:2020yrt,Mastrototaro:2021wzl,Boyarsky:2021yoh}. However, such studies have intrinsic limitations. This is because the computational complexity of the discretization approach quickly increases with the maximal energy of the injected neutrinos,\footnote{Assuming the linear grid in momentum space, which is the only adequate model-agnostic choice, the scaling of the computational time is $t \propto E_{\nu}^{\alpha+1+1}$, with $\alpha\geq 2$ coming from the irreducibility of the integration of the collision integral, the power of one from the number of equations, and the remaining power from the scaling of the timestep required to resolve the interactions.} requiring days or weeks to solve for GeV-scale LLPs that inject high-energy neutrinos~\cite{Sabti:2020yrt}. The complexity of the discretization method forced people to adopt simplified descriptions of the neutrino dynamics~\cite{Fradette:2017sdd,Escudero:2018mvt,EscuderoAbenza:2020cmq,Gelmini:2020ekg,Chao:2022mcu,Chen:2024cla,Chang:2024mvg}, which inaccurately describes their evolution unless neutrinos are completely decoupled or specific assumptions about the underlying new physics take place, such as thermality of neutrino injection.

\emph{Second}, various implementations of the discretization approach yield qualitatively contradictory results. For instance, considering LLPs decaying into high-energy neutrinos at plasma temperature $T\simeq 1\text{ MeV}$, Refs.~\cite{Dolgov:2000jw,Mastrototaro:2021wzl} predict an increase in $N_{\text{eff}}$, whereas the works~\cite{Boyarsky:2021yoh,Rasmussen:2021kbf} show, counterintuitively, its decrease. The origin of the discrepancy is not known, as the methods are very complicated, and some of the codes are not publicly accessible (see also a discussion in~\cite{Boyarsky:2021yoh}).

\emph{Third}, recent theoretical advances~\cite{Akita:2024ork,Akita:2024nam} show important effects of the evolution of secondary decay products of LLPs, such as mesons and muons, that have not been accounted for in the previous studies. These particles do not just lose kinetic energy and decay at rest but may disappear due to self-annihilations and interactions with nucleons, which completely change their impact on neutrinos.

\emph{Fourth}, even with these features considered, most previous studies examined the impact of new physics assuming the standard cosmic setup, such as the absence of neutrino asymmetry and dark radiation. While this assumption is natural from the point of simplicity, it may actually lead to a wrong conclusion that the given parameter space of new physics is inevitably ruled out by cosmological observations. Varying the setup can enhance or relax the constraints~\cite{Gelmini:2020ekg}, which is very important in light of the complementarity with laboratory searches, as is sketched in Figure~\ref{fig:hnl-complementarity}.

\textbf{Neutrino DSMC.} To overcome the limitations of existing methods, we introduce a novel approach based on the Direct Simulation Monte Carlo (DSMC) method~\cite{bird1970direct,ivanov1991theoretical,stefanov2019basic}, significantly adapted for studying neutrino thermalization in the early Universe. Details are provided in the accompanying paper~\cite{Ovchynnikov:2024rfu}, and here we summarize the main points. 

DSMC is a numerical technique traditionally used to solve the Boltzmann equation for gases with short-range interactions by simulating the motion and collisions of a large number of particles. The evolution of the system is simulated over discrete time steps $\Delta t$, during which particles undergo free streaming, the system is split into sub-volumes called \emph{cells}, and within each cell, particles' interactions are accepted probabilistically based on the largeness of their interaction cross-section and then simulated if accepted. 

In our neutrino DSMC approach, we represent the neutrino distribution function by a large ensemble of particles, each characterized by its energy and lepton charge. Several key adaptations are necessary to apply DSMC to the Early Universe plasma. First, the \emph{expansion of the Universe} is incorporated by redshifting particles' energies and the system's volume at each time step according to the Hubble parameter $H$.

Second, the \emph{rapid equilibration of the electromagnetic (EM) plasma} allows us to treat the EM sector as a thermal bath characterized by a temperature $T$. Instead of simulating individual electrons and positrons, we sample their energies from equilibrium Fermi-Dirac distribution at temperature $T$, which is updated inside each cell based on the energy exchange with neutrinos.

Third, \emph{quantum statistics} is included by accounting for Pauli blocking factors in the collision probabilities, at the last step of the probabilistic acceptance of the interaction.

Fourth, \emph{neutrino oscillations} are incorporated by allowing neutrinos to change flavor between time steps, according to energy-dependent oscillation probabilities averaged over the time and energy scales relevant for the Early Universe~\cite{Ruchayskiy:2012si,Sabti:2020yrt}.

Finally, \emph{decaying particles} (LLPs) are included by simulating their decay processes and subsequently adding the decay products into the DSMC particles list, according to the exponential law governing the decay. The phase space of decay products can be obtained using existing Monte Carlo tools like \texttt{PYTHIA8}~\cite{Bierlich:2022pfr} or \texttt{SensCalc}~\cite{Ovchynnikov:2023cry}, ensuring model independence and flexibility. This allows us to handle complex decay chains and high-mass particles without the need for analytical expressions for the matrix elements.

There are numerous advantages of our neutrino DSMC approach. The method is \emph{model-agnostic}, capable of handling arbitrary interaction processes without requiring analytic expressions for collision integrals or the phase space of the interactions, with computational time scaling linearly with the maximal neutrino energies involved. The scaling is unavoidable and follows from the behavior of the timestep required to resolve the high-energy neutrino interactions.\footnote{In the case of extremely large neutrino energies involved $E_{\nu}\gtrsim 10\text{ GeV}$, we would need to increase the number of computational particles to properly describe thermalization with DSMC. However, the resulting dependence becomes linear, and this worst-case total scaling is quadratic -- still much better than the (at least) quartic power scaling in the state-of-the-art approaches (see Appendix B of our companion paper~\cite{Ovchynnikov:2024rfu}).} This efficiency allows us to simulate scenarios with high-energy neutrinos (up to GeV scales) where traditional discretization methods become impractical. Moreover, the method is conceptually straightforward, inherently conserving energy and avoiding stability issues associated with the discretization schemes. Finally, it is flexible: new interactions, particles, and variations of the cosmic setup may be easily incorporated.

We have developed a prototype code including modules for simulating particle interactions, decay processes of LLPs, and the evolution of the EM plasma.\footnote{The prototype code may be provided upon request and will be made public once the full \texttt{C++} implementation is ready.} We have validated our approach by comparing its predictions with existing methods~\cite{EscuderoAbenza:2020cmq,Akita:2020szl,Boyarsky:2021yoh} in well-understood scenarios, such as thermalization of neutrino and EM species having initial thermal spectra and the evolution of neutrino spectra under instant non-thermal neutrino injections. 

The performance of our DSMC implementation exceeds one of the traditional methods already for maximal neutrino energies $\simeq 50-100\,\text{MeV}$. In the domain of higher energies, our algorithm quickly becomes more efficient by orders of magnitude. This is because the running time of our approach scales just linearly, which is due to the behavior of the timestep, whereas the discretization approaches have a much worse scaling. This demonstrates the potential of our approach to handle complex scenarios involving high-energy neutrinos. Moreover, the code's performance will be substantially improved if switching to solely \texttt{C++} implementation.

\textbf{Case studies.} Given the flexibility of Monte-Carlo simulations and the efficiency of the DSMC scheme, it is model-agnostic. It is possible to incorporate various processes such as neutrino self-interactions, decays of LLPs, or vary cosmic setup, as well as consider any parameter space, particularly not being restricted to small neutrino energies. This is a subject of numerous future studies. Here, we will concentrate on a few examples of decaying LLPs, demonstrating their capabilities. For definiteness, we assume the lifetime $\tau = 0.03\text{ s}$, which is close to the edge of what may be probed with CMB and BBN, and fix the initial LLP energy density by requiring $\rho_{\rm LLP}/\rho_{\rm \gamma,\nu,e^{\pm}} = 0.2$ at $T = 5\text{ MeV}$. We also do not include neutrino oscillations to make the discussion more transparent. For the range of the LLP masses $m$, we include the GeV scale, which is extremely complicated to explore with state-of-the-art methods due to the computational inefficiency. 

\begin{figure}[t!]
    \centering
    \includegraphics[width=0.9\linewidth]{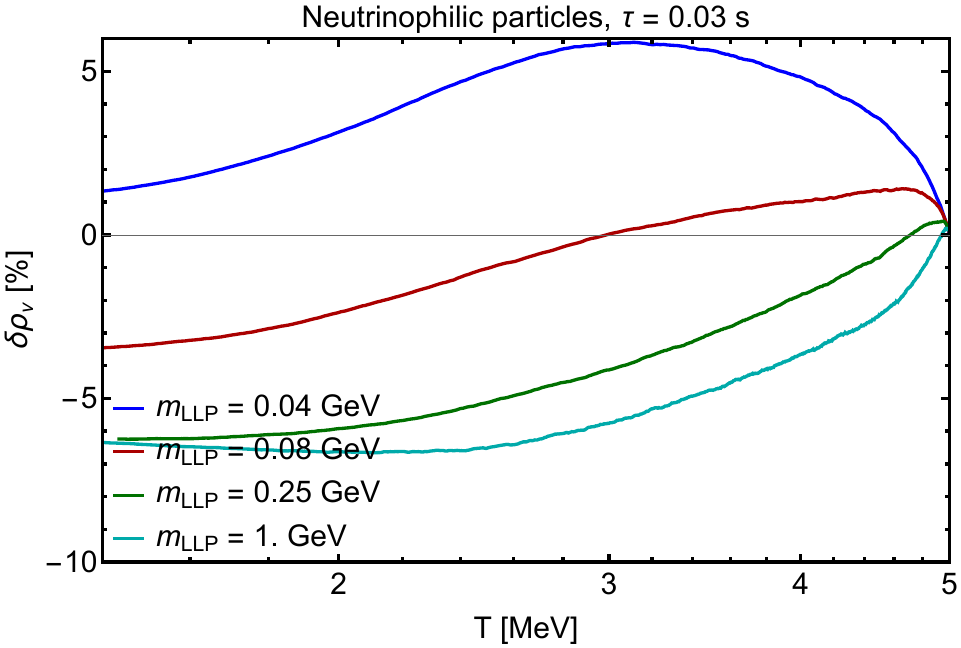}
    \caption{Application of our approach to the case of Long-Lived Particles (LLPs) with different masses $m$ and lifetime $\tau = 0.03\text{ s}$ decaying solely into a pair of neutrinos in the MeV plasma. The figure shows the evolution of the quantity $\delta \rho$, defined by Eq.~\eqref{eq:delta-rho}, with the temperature of the electromagnetic plasma $T$. $\delta \rho$ is intimately related to the correction to the effective number of neutrino species $\Delta N_{\rm eff} = N_{\rm eff} - N_{\rm eff}^{\Lambda\text{CDM}}$. From the plot, we see that for large masses $m\gg 3T$, $N_{\rm eff}$ decreases; this is true for a genetic heavy LLP that has other decay modes into SM particles. This finding resolves the discrepancy between existing studies~\cite{Dolgov:2000jw,Boyarsky:2021yoh,Mastrototaro:2021wzl,Rasmussen:2021kbf}, ambiguously predicting the sign of $\Delta N_{\rm eff}$.}
    \label{fig:neutrinophilic}
\end{figure}

\emph{The first case} is the LLPs decaying solely into a neutrino-antineutrino pair, each with energy $E_{\nu} = m/2$, in the thermal SM plasma. Examples of such scenarios are majorons and other neutrinophilic mediators~\cite{Chikashige:1980qk,Chikashige:1980ui,Kelly:2019wow,Akita:2023qiz}. Depending on $m$, $E_{\nu}$ may be much higher than the typical thermal neutrino energy $3T$, where $T$ is the plasma temperature. Fig.~\ref{fig:neutrinophilic} shows the dynamics of the quantity 
\begin{equation}
\delta \rho_{\nu}(T) \equiv \left(\frac{\rho_{\rm EM}}{\rho_{\nu}}\right)_{\Lambda \text{CDM}}\frac{\rho_{\nu}}{\rho_{\rm EM}} - 1,
\label{eq:delta-rho}
\end{equation}
where $\rho_{\nu},\rho_{\rm EM}$ denote the energy densities of neutrinos and EM particles, and $(\rho_{\nu}/\rho_{\rm EM})_{\Lambda \text{CDM}}$ is the ratio of the energy densities in the standard cosmological model. $\delta \rho_{\nu}$ evolves due to the decays of LLPs and is the dynamical analog of $N_{\rm eff}$. In particular, its sign is uniquely related to the sign of the correction of $N_{\rm eff}$ with respect to its value in $\Lambda$CDM. 

As the initial injection is solely to neutrinos, each decay initially pumps energy to the neutrino sector, and naively, $\delta \rho_{\nu}$ must be positive throughout the evolution. However, depending on $m$, the subsequent thermalization of neutrinos leads to the transition to negative values and hence to a decrease in $N_{\rm eff}$. The reason is that the energy transfer rates between the neutrinos and the EM plasma grow with the energies of interacting particles. By constantly injecting non-thermal neutrinos, we shift the balance of the exchange to the EM plasma. The latter thermalizes instantly, and hence the backward energy transfer is not enhanced. The microscopic of this thermalization is discussed in detail in the companion paper~\cite{Ovchynnikov:2024rfu} using the toy scenario of instant neutrino injections. 

The same result qualitatively holds for the LLPs decaying into muons and pions, such as Higgs-like scalars, axion-like particles, neutralinos, and vector mediators coupled to quark currents. The reason is that $\mu$s and $\pi$s decay into high-energy neutrinos, so their energy ends up being in the EM sector. 

As the neutrino DSMC is a completely independent approach to solving the neutrino Boltzmann equation, we close the mentioned discrepancy between various Boltzmann solvers, supporting the results of~\cite{Boyarsky:2021yoh,Rasmussen:2021kbf}. This finding allows us to conclude that generic heavy LLPs decaying into SM species (including the whole range of models mentioned above) at MeV temperatures in the standard cosmological setups unavoidably decrease $N_{\rm eff}$, which will be very important in light of interpreting the results of upcoming CMB measurements.

\begin{figure}[t!]
    \centering
    \includegraphics[width=0.95\linewidth]{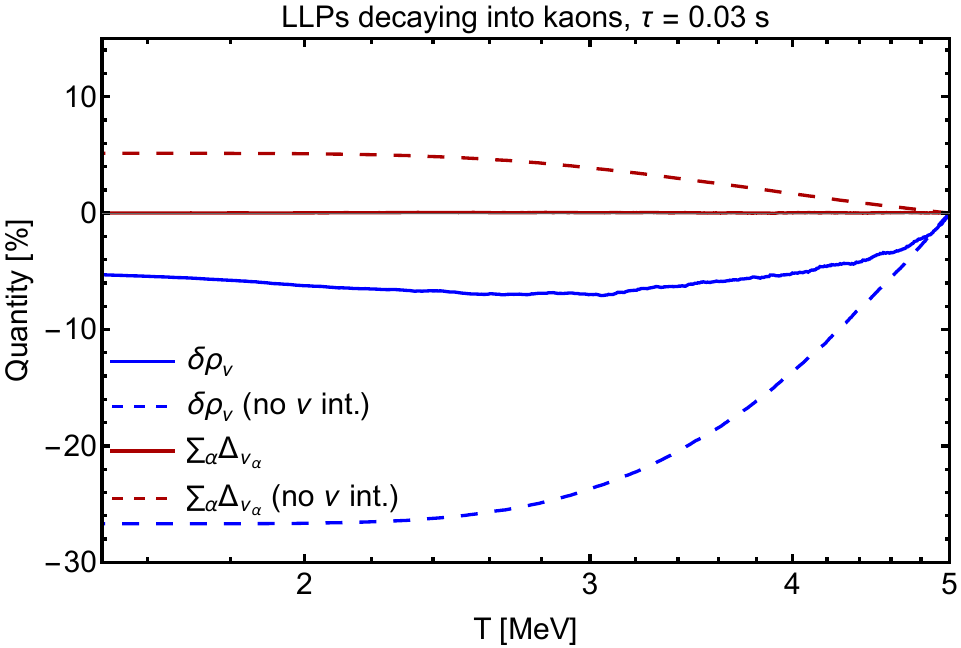}
    \caption{Application of the DSMC method to the LLPs decaying exclusively into pairs of charged kaons. The plot shows the neutrino energy density perturbation, $\delta\rho_{\nu}(T)$, along with the neutrino-antineutrino energy density asymmetries, $\Delta_{\nu_{\alpha}}$ (see Eq.~\eqref{eq:neutrino-asymmetry}). These asymmetries arise from the different behaviors of $K^{-}$ and $K^{+}$ in the primordial plasma; to show its potential magnitude, we indicate the development of the asymmetry while turning off neutrino interactions. However, when including neutrino interactions, we find that they instantly erase this asymmetry, thanks to the efficient neutrino-antineutrino annihilations at such high temperatures.}
    \label{fig:into-kaons}
\end{figure}

\emph{The second scenario} we consider is a more complicated setup where an LLP decays into a pair of charged kaons~\cite{Pospelov:2010cw}. Such a scenario is realized in various models, including Higgs-like scalars~\cite{Boiarska:2019jym} and generic mediators coupled to the quark currents, such as dark photons and $B-L$ mediators~\cite{Ilten:2018crw} (see also~\cite{Alekhin:2015byh,Beacham:2019nyx,Antel:2023hkf} for a more generic discussion). They resonantly mix with scalar or vector mesons such as $f_{0},\phi$, that decay into kaons. 

The case is special and even more complex, as the dynamics of kaons in the primordial plasma are non-trivial and not described just by decays-at-rest (see the recent studies~\cite{Akita:2024nam,Akita:2024ork} for detail). Namely, $K^{+}K^{-}$ may efficiently annihilate with themselves. More importantly, $K^{-}$ evolves differently compared to $K^{+}$: it intensively participates in the interactions with nucleons~\cite{Reno:1987qw}. At large temperatures, the process may completely dominate over decays. This means that the decay probability of $K^{-}$ in the primordial plasma is significantly lower than that of $K^{+}$. This difference induces further asymmetries in the evolutions of muons and pions produced by their decays and scatterings. As $K^{-}$s decay into neutrinos with energies $E_{\nu}\simeq m_{K}/2$ while pions and muons into much less energetic $\nu$s, this dynamics may result in non-trivial effects, including the neutrino-antineutrino energy distribution asymmetry.

Due to the computational complexity in this setup, the authors of~\cite{Akita:2024nam,Akita:2024ork} have left a detailed investigation of the role of this asymmetry for future work. DSMC is a perfect tool to explore such models. Therefore, in addition to the evolution of $\delta \rho_{\nu}$, we are also interested in the development of the energy asymmetries between neutrinos and antineutrinos, which we define as
\begin{equation}
    \Delta_{\nu_{\alpha}} = \frac{\rho_{\nu_{\alpha}}-\rho_{\bar{\nu}_{\alpha}}}{\rho_{\nu_{\alpha}}+\rho_{\bar{\nu}_{\alpha}}}
    \label{eq:neutrino-asymmetry}
\end{equation}
To trace it, we used the output of the public code provided by Refs.~\cite{Akita:2024nam,Akita:2024ork} and, for the first time, fully incorporated the dynamics of the kaons $K$ in the evolution of neutrinos using our approach. 

We show the evolution of $\delta \rho_{\nu}$ and $\Delta_{\nu_{\alpha}}$ in Fig.~\ref{fig:into-kaons}. Despite the initial $\Delta_{\nu_{\alpha}}\neq 0$ induced by asymmetric decays, the efficient neutrino-neutrino interactions turn it to zero. The situation may change for higher lifetimes, as the interactions are less efficient, and partial asymmetry will survive.  

\textbf{Conclusions.} In light of the significant computational challenges, recent theoretical advances, and forthcoming cosmological observations and laboratory searches for new physics, there is a clear need for an efficient and model-independent approach to assessing the impact of new physics on primordial neutrinos. We have presented such an approach based on the Direct Simulation Monte Carlo (DSMC) method, which effectively addresses these requirements. While our current implementation serves as a proof-of-concept, it already demonstrates the capability to make meaningful predictions and resolve existing discrepancies in the literature. 

Moving forward, we plan to apply this versatile method to a broad range of new physics models affecting cosmic neutrinos. This will provide valuable insights for both the cosmology and particle physics communities, enhancing our understanding of the Early Universe and defining the target parameter space for future laboratory searches for new physics. We will go beyond the minimalistic assumptions on the cosmic setup used in past studies and compute the uncertainty range of the cosmological bounds. Another promising direction to explore is the application of the DMSC method to study the dynamics of supernova explosions in the presence of new physics.

\textbf{Acknowledgements.} We thank Stefan Stefanov for reviewing our DSMC implementation, Fabio Peano, Luís Oliveira e Silva, and Kyrylo Bondarenko for helpful discussions, and Gideon Baur and Miguel Escudero for reading the manuscript. M.O. acknowledges support from the European Union's Horizon 2020 research and innovation program under the Marie Sklodowska-Curie grant agreement No. 860881-HIDDeN.

\bibliography{main.bib}

\end{document}